\documentclass[%
 preprint,
showpacs,preprintnumbers,
 amsmath,amssymb,
 aps,
]{revtex4-1}

\usepackage{graphicx}

\newcommand{\myfig}[4][ht]{
\begin{figure}[#1]
\centering
\includegraphics[#2]{#3}
\caption{#4\label{#3}}
\end{figure}
}

\graphicspath{{images/}}

\usepackage{dcolumn}
\usepackage{bm}

\begin{document}

\title{Spatiotemporal flux memory in nondiffusive transport}


\author{Bjorn Vermeersch}
\email{bvermeer@purdue.edu}
\author{Ali Shakouri}
\email{shakouri@purdue.edu}
\affiliation{\vspace{3mm}Birck Nanotechnology Center, Purdue University, West Lafayette, Indiana 47907, USA}
\date{\today}

\begin{abstract}
Anomalous diffusion constitutes a relation between tracer flux and tracer density gradient that is inherently nonlocal in space and/or time. Previous studies emphasize the non-Gaussian character of the tracer distribution that arises from adjusted constitutive relations but did not investigate the flux-gradient memory itself. Here, we present a universal analytic framework that enables systematic characterisation of nonlocality in a wide variety of transport regimes. A generalised diffusivity kernel $D^{\ast}$ fully embodies the spatiotemporal flux memory with respect to the gradient. An extension of the flux-gradient relation for subdiffusive transport is also proposed. Several conservation and invariance properties can be deduced, including that Poissonian flight processes have no flux memory in time while fractional time diffusion has no flux memory in space. We derive analytical expressions for $D^{\ast}(x,t)$ in several types of anomalous transport dynamics that are commonly encountered in practice, being fractional diffusion equations, tempered L\'evy superdiffusion, and tempered fractional time diffusion. This detailed knowledge of the shape and nature of the flux memory, and the corresponding length and time scales over which nonlocal effects are physically important, remain completely hidden in conventional analyses based on tracer distributions or flux-gradient diagrams. Practical capabilities include the interpretation of microscale heat superdiffusion experiments. Overall, the theory can serve as a valuable framework for anomalous transport dynamics across multiple disciplines.
\end{abstract}
\maketitle
\section{INTRODUCTION}
Transport phenomena in which the mean square displacement (MSD) evolves in time as $\sigma^2(t) \sim t^{\gamma}$ with $\gamma \neq 1$ are designated as `anomalous'. Some processes have mathematically divergent MSDs, and yet others exhibit $\gamma = 1$ while their underlying kinetics clearly deviate from regular diffusion. This work deals with examples of each of those cases, and we will employ `anomalous' in the broad sense to denote any transport dynamics that cannot be described by regular diffusion equations with stationary and homogeneous parameters.
\par
Phenomena of this nature have been encountered in a wide variety of disciplines. Two comprehensive reviews by Metzler and Klafter \cite{klafter1,klafter2} and the references therein give a good overview. Here, we briefly mention a by no means exhaustive list of examples in thermal physics \cite{heatconduction-nonlocal1,heatconduction-nonlocal2,heatconduction-ballisticdiffusive,heatconduction-nanoparticles}, turbulent fluid and plasma dynamics \cite{turbulence-stretchedexp,plasma-electrontransport,plasma-fractionaldiffeq,plasma-heat-stretchedexp,plasma-heatwave,plasmaturbulence}, chemistry \cite{chemistry-levy1,chemistry-levy2}, hydrology and geophysics \cite{geophysics-review,geophysics-temperedsubdiffusion2,geophysics-temperedsubdiffusion1,geophysics-temperedlevy,hydrology-temperedlevy}, biology and medicine \cite{biology-animals,biology-DNA,medicine-MRI1,medicine-MRI2,medicine-drugrelease}, and finance \cite{finance-mantegna,finance-temperedlevy1,finance-temperedlevy2}. 
\par
In regular diffusive transport, the tracer flux $\vec{q}(\vec{r},t)$ only depends on the local and instantaneous tracer density gradient $\vec{\nabla} P(\vec{r},t)$. In one-dimensional configurations we have
\begin{equation}
q(x,t) = -D_0 \frac{\partial P}{\partial x} (x,t) \label{FGR0}
\end{equation}
where the proportionality factor $D_0$ is the diffusivity (unit m$^2$/s) of the transport medium. In mass transfer applications, this flux-gradient relation (FGR) is known as Fick's law, relating the mass flux to the concentration gradient. For thermal transport, the FGR is usually written as Fourier's law $q = - \kappa_0 \, \partial_x T$ that relates the heat flux to the temperature gradient through the thermal conductivity $\kappa_0$. In the equivalent formulation (\ref{FGR0}), $P$ denotes the thermal energy density $C_0 T$ where $C_0$ is the volumetric heat capacity with $D_0 = \kappa_0/C_0$.
\par
At length and/or time scales comparable to the mean free paths and/or relaxation times of the microscopic heat/mass carriers, (\ref{FGR0}) breaks down, and anomalous transport emerges. Such regimes must inherently constitute a nonlocal FGR \cite{heatconduction-nonlocal1}, meaning that the flux at a given place and time is codependent on the density gradient at other locations and/or earlier times. One can equivalently say that the FGR develops spatial and/or temporal memory. Earlier literature on nondiffusive transport acknowledges \cite{chemistry-levy2,fractionalfick-nonlocal,geophysics-temperedsubdiffusion2,heatconduction-nanoparticles,philipallen,plasma-electrontransport,plasma-fractionaldiffeq,plasma-heat-stretchedexp} or explicitly addresses \cite{heatconduction-ballisticdiffusive,heatconduction-nonlocal1,heatconduction-nonlocal2,koh-nonlocal,plasma-heatwave} FGR delocalisation but did not pursue characterisation of the associated memory aspects themselves. Instead, adjustments to constitutive relations are used as a means to derive the tracer distribution and emphasize its anomalous (non-Gaussian) character. Analyses commonly involve elaborate numerical schemes and are scattered across several disciplines, thereby defying easy distillation of generic trends. Some works \cite{fractionalfick-nonlocal} explored the transport dynamics through flux-gradient diagrams which provide snapshot comparisons of the flux to the local gradient by plotting $q(x,t_0)$ versus $- \partial_x P(x,t_0)$ using $x$ as a parameter. Although the shape and curvature of such diagrams can serve as rough indication of transport anomaly, this approach still doesn't provide much insight into the precise nature of the underlying nonlocal effects.
\par
In this work, we develop a unified framework for describing and characterising FGR memory that is applicable to a wide variety of anomalous transport regimes. Prominent attention is given to analytical derivation of universal properties in order to facilitate broad insights. Key findings include closed form expressions for the FGR memory kernel in several archetypical nondiffusive transport dynamics that are commonly encountered in practical applications. In the process, we also determine the length and time scales over which flux memory effects are physically important, thereby answering the key question that naturally arises when a given transport regime becomes delocalised.
\section{THEORETICAL FRAMEWORK}
We assume a homogeneous, isotropic medium and perform the discussion, as is customary, in one spatial dimension. This only poses minor practical limitations since the essential physics are usually dominated by either in-plane or cross-plane transport. We seek to generalise the conventional Fourier/Fick law (\ref{FGR0}) to nondiffusive transport regimes in the form
\begin{equation}
q(x,t) = - \int \limits_{0}^{t} \mathrm{d}t' \int \limits_{-\infty}^{\infty} \mathrm{d}x' D^{\ast}(x-x',t-t') \frac{\partial P}{\partial x}(x',t') \label{convolution}
\end{equation}
Here, $D^{\ast}(x',t')$ is a generalised diffusivity kernel that embodies the spatial and temporal memory of the heat/mass flux with respect to the energy/concentration gradient. This quantity must not be confused with the `effective' diffusivity that arises when interpreting the anomalous dynamics phenomenologically through $q(x,t) = -D_{\text{eff}}(x,t) \, \partial_x P(x,t)$ as is commonplace in microscale heat conduction experiments \cite{cahill,siemens,minnich,johnson,malen}.
\par
For reasons that will become clear in section IV.A, the intuitive form (\ref{convolution}) fails to describe transport regimes with fractal space dimension $\mathbb{D}_x$ and/or fractal time dimension $\mathbb{D}_t$ smaller than one. Equation (\ref{convolution}) can be extended to its final universal form
\begin{equation}
\left[ \frac{\partial}{\partial x} \right]^{\chi} q(x,t) = - \left[ \frac{\partial}{\partial t} \right]^{\eta} \int \limits_{0}^{t} \mathrm{d}t' \int \limits_{- \infty}^{\infty} \mathrm{d}x' D^{\ast}(x-x',t-t') \frac{\partial P}{\partial x}(x',t') \label{convolution2}
\end{equation}
where we have introduced the binary specifiers $(\chi,\eta)$ defined as
\begin{equation}
\chi = \begin{cases} 1 \text{ if } 0 < \mathbb{D}_x < 1 \\ 0 \text{ if } 1 \leq \mathbb{D}_x \leq 2 \end{cases} , \quad \eta = \begin{cases} 1 \text{ if } 0 < \mathbb{D}_t < 1 \\ 0 \text{ if } 1 \leq \mathbb{D}_t \leq 2 \end{cases}
\end{equation}
\par
After spatial Fourier ($x \leftrightarrow \xi$) and temporal Laplace ($t \leftrightarrow s$) transformations the double convolution and linear operators in (\ref{convolution2}) simplify to
\begin{equation}
(j\xi)^{\chi} \, q(\xi,s) = - s^{\eta} \, D^{\ast}(\xi,s) \, [j \xi P(\xi,s)] \label{qP1}
\end{equation}
where $j$ denotes the complex unit. Conservation of energy/mass imposes that $- \partial_x q = \partial_t P$ at all places and times, which in transformed variables reads
\begin{equation}
-j \xi q(\xi,s) = s P(\xi,s) - P(\xi,t=0) \label{qP2}
\end{equation}
Let us now consider a planar source located inside the medium at $x=0$ that injects one unit of heat/mass per m$^2$ at time zero. This provides the fundamental single pulse response of the system from which all dynamic properties can be derived. Given the initial condition $P(x,t=0) = \delta(x)$ we have $P(\xi,t=0) = 1$, so combining (\ref{qP1}) and (\ref{qP2}) produces
\begin{equation}
D^{\ast}(\xi,s) = \frac{j^{\chi}}{s^{\eta} \, \xi^{2-\chi}} \left[ \frac{1}{P(\xi,s)} - s \right] \label{kappamacro}
\end{equation}
Since our derivation solely relied on elementary conservation laws and basic isotropy assumptions, the result (\ref{kappamacro}) is universally applicable to a broad array of anomalous diffusion dynamics. Before exploring specific examples in closer detail, we verify that the framework correctly recovers to a localised FGR for regular diffusion. This process has fractal dimensions $(\mathbb{D}_x,\mathbb{D}_t) = (2,1) \Rightarrow (\chi,\eta) = (0,0)$ and is characterised by a Gaussian single pulse response $P(\xi,s) = (s + D_0 \xi^2)^{-1}$. Eq. (\ref{kappamacro}) produces $D^{\ast}(\xi,s) \equiv D_0$ and therefore $D^{\ast}(x',t') = D_0 \delta(x')\delta(t')$, so the convolution (\ref{convolution}) indeed reduces to the memoryless Fourier/Fick law as appropriate.
\section{GENERIC FLUX MEMORY PROPERTIES}
Many transport phenomena can be understood in terms of random motion of tracer particles. We briefly review some key fundamentals following \cite{klafter1} and \cite{CTRWmaster}. In the stochastic context, the single pulse response $P(x,t)$ is the probability density function associated with the chance to find a randomly wandering tracer in location $x$ at time $t$ after it was released by the source in $x=0$ at $t=0$. The tracer performs consecutive independent transitions that move it by a distance $\zeta$ in time span $\vartheta$ randomly chosen from a joint distribution $\Phi(\zeta,\vartheta)$. For clarity, we note that $\Phi$ is often called the `memory function' of the stochastic process and must not be confused with the `FGR memory' $D^{\ast}(x',t')$ investigated here. Flight processes constitute an important subgroup and have as defining feature that the jump length is stochastically independent from the waiting time: $\Phi(\zeta, \vartheta) = \phi(\zeta) \cdot \varphi(\vartheta)$. This decoupling simplifies the mathematical description and is physically justified at length and time scales over which the finite carrier propagation velocity does not restrict the transport, which is adequate for most practical applications. The Montroll-Weiss equation \cite{CTRWmaster} provides a closed form solution for the Fourier-Laplace single pulse response $P(\xi,s)$ induced by the flight process in terms of the transformed distributions $\phi(\xi)$ and $\varphi(s)$:
\begin{equation}
P(\xi,s) = \frac{1 - \varphi(s)}{s [1 - \phi(\xi) \varphi(s)]} = \frac{\Psi(s)}{s \left[ \Psi(s) + \psi(\xi) \right]}\label{CTRWmaster}
\end{equation}
where we introduced $\psi(\xi) = 1-\phi(\xi)$ and $\Psi(s) = 1/\varphi(s) - 1$ for notation convenience. We consequently obtain
\begin{equation}
D^{\ast}(\xi,s) = j^{\chi} \, \frac{\psi(\xi) / \xi^{2-\chi}}{\Psi(s)/s^{1-\eta}} \label{kappaflights}
\end{equation}
We now take a closer look at two types of nondiffusive transport which each exhibit a powerful invariance in their FGR memory.
\subsection{Temporal invariance in Poissonian flight processes}
Let us consider transport in which the carrier wait time is exponentiallly distributed with average $\left< \vartheta \right> = \tau$. The occurrence of jumps over time then follows a Poisson process, with average event frequency $1/\tau$. With $\Psi(s) = s\tau$ , the single pulse response (\ref{CTRWmaster}) takes the form
\begin{equation}
P(\xi,s) = \frac{1}{s + \psi(\xi)/\tau} \label{Ppoisson}
\end{equation}
In real space and time domain, the tracer density and associated flux are then given by
\begin{eqnarray}
P(x,t) & = & \frac{1}{2\pi} \int \limits_{-\infty}^{\infty} \exp \left[ - \psi(\xi) \, \frac{t}{\tau} \right] \, \exp(j\xi |x|) \, \mathrm{d}\xi \label{Pxtpoisson} \\
q_{\pm}(x,t) & = & \frac{\pm 1}{2\pi} \int \limits_{-\infty}^{\infty} \frac{\psi(\xi)}{j \xi \tau} \exp \left[ - \psi(\xi) \, \frac{t}{\tau} \right] \, \exp(j\xi |x|) \, \mathrm{d}\xi \label{qxtpoisson}
\end{eqnarray} 
with the flux specifier $\pm$ denoting the sign of $x$. Poissonian processes have $\mathbb{D}_t = 1$ ($\eta = 0$), and (\ref{kappamacro}) reduces to
\begin{equation}
D^{\ast}(\xi) = \frac{j^{\chi} \, \psi(\xi)}{\tau \, \xi^{2-\chi}} \label{kappapoisson}
\end{equation}
The $s$ dependence has vanished from the generalised diffusivity: \textit{the FGR in Poissonian transport regimes has no temporal memory}. The reasoning can also be made in reverse, showing that a transport phenomenon that lacks temporal flux memory is inherently governed by Poissonian flight dynamics. L\'evy superdiffusion forms a specific example. In this case we have $\psi(\xi) = L^{\alpha} |\xi|^{\alpha}$ with $1 \leq \alpha < 2$ the governing exponent and $L$ a characteristic length. The jump length distribution no longer decays exponentially as in regular diffusion but exhibits `heavy tails' $\phi(\zeta \gg L) \sim |\zeta|^{-(\alpha+1)}$. As a result, L\'evy carriers do not perform Brownian motion but instead generate clustered patterns with fractal space dimension $\mathbb{D}_x = \alpha$ \cite{klafter1}. The associated single pulse response obeys an alpha-stable distribution $P(\xi,s) = 1/(s + D_{\alpha} \xi)$ where $D_{\alpha} = L^{\alpha}/\tau$ is the so called fractional diffusivity (unit m$^{\alpha}$/s). We consequently find $D^{\ast}(\xi) = D_{\alpha}/|\xi|^{2-\alpha}$, so the FGR memory decays algebraically in space: $D^{\ast}(x') \sim D_{\alpha} \, |x'|^{-(\alpha-1)}$.
\par
An interesting conservation property arises if the carrier jump length distribution that drives the Poissonian process has a finite second moment $\left< \zeta^2 \right> = 2 L^2$. These processes become diffusive at long length and time scales with `asymptotic' diffusivity $D_{\infty} = L^2/\tau$, and accounting for $\psi(\xi \rightarrow 0) \simeq L^2 \xi^2$ in (\ref{kappapoisson}) produces
\begin{equation}
\mathbb{D}_x \geq 1 : \quad D^{\ast}(\xi \rightarrow 0) = D_{\infty} \quad \leftrightarrow \quad \int \limits_{-\infty}^{\infty} D^{\ast}(x') \mathrm{d}x' = D_{\infty} \label{Poissonconservation}
\end{equation}
\textit{The total spatial flux memory content in Poissonian transport with long-term diffusion recovery equals the asymptotic diffusivity at all times.} This offers an insightful perspective on the FGR delocalisation: the overall generalised diffusivity `budget' is fixed by the long-term property $D_{\infty}$ but the anomalous short-term dynamics determine how it gets distributed across space. A commonly encountered Poissonian process with diffusive recovery is given by tempered L\'evy transport, which we investigate in detail in section IV.B.
\subsection{Spatial invariance in fractional time diffusion}
Let us consider transport in which the carrier jump length distribution obeys $\psi(\xi) = L^2 \xi^2$ at all spatial frequencies just like in regular diffusion, with $L$ again denoting a characteristic length. Now the single pulse response (\ref{CTRWmaster}) takes the form
\begin{equation}
P(\xi,s) = \frac{\Psi(s)}{s [\Psi(s) + L^2 \xi^2]} \label{PBrownian}
\end{equation}
For the real space and time domain counterpart and associated flux we have
\begin{eqnarray}
P(x,t) & = & \mathcal{L}^{-1} \left[ \frac{\sqrt{\Psi(s)}}{2 s L} \exp \left( - \sqrt{\Psi(s)} \, \frac{|x|}{L} \right) \right] \label{PxtBrownian}\\
q_{\pm}(x,t) & = & \mathcal{L}^{-1} \left[ \pm \frac{1}{2} \exp \left( - \sqrt{\Psi(s)} \, \frac{|x|}{L} \right) \right] \label{qxtBrownian}
\end{eqnarray}
In the spirit of earlier works \cite{fractionaltimediffusion}, we will refer to transport regimes of this type as `fractional time diffusion' since the process preserves the spatial dimension $\mathbb{D}_x = 2$ characteristic of Brownian motion but permits non-exponential wait time distributions. These transport regimes always have a well defined MSD:
\begin{equation}
\sigma^2(s) = - \frac{\partial^2 P(\xi,s)}{\partial \xi^2} \biggr|_{\xi=0} = \frac{2L^2}{s \Psi(s)} \label{sigmasquareBrownian}
\end{equation}
Given (\ref{PBrownian}) as functional form, (\ref{kappamacro}) reduces to
\begin{equation}
D^{\ast}(s) = \frac{s^{1-\eta} L^2}{\Psi(s)} \label{kappabrownian}
\end{equation}
The $\xi$ dependence has vanished from the generalised diffusivity: \textit{the FGR in fractional time diffusion has no spatial memory}. Here, too, the condition is both necessary and sufficient: transport processes that lack flux memory in space must be governed by fractional time diffusion dynamics. Notice that the temporal flux memory is closely related to the MSD evolution, since $D^{\ast}(s) = s^{2-\eta} \, \sigma^2(s) / 2$ per (\ref{sigmasquareBrownian}) and (\ref{kappabrownian}). A specific example of fractional time diffusion is given by what we will term `Mittag-Leffler subdiffusion'. Here, $\Psi(s) = (s \tau)^{\beta}$ with $\tau$ a characteristic time scale and $0 < \beta < 1$ the governing exponent. The associated wait time has a Mittag-Leffler distribution \cite{MLdistribution} given by
\begin{equation}
\varphi(\vartheta) = \mathcal{L}^{-1} \left[ \frac{1}{1 + (s\tau)^\beta} \right] = -\frac{\beta}{\tau} \sum \limits_{n=1}^{\infty} \frac{(-1)^n \, n \, (\vartheta/\tau)^{\beta n - 1}}{\Gamma(\beta n + 1)} \label{mittagleffler}
\end{equation}
which has a heavy tail $\varphi(\vartheta \gg \tau) \sim \vartheta^{-(\beta+1)}$. Given that the process has $\mathbb{D}_t = \beta$ and therefore $\eta = 1$, we obtain $D^{\ast}(s) = D_{\beta}/s^{\beta}$, with $D_{\beta} = L^2/\tau^{\beta}$ the fractional diffusivity constant (unit m$^2$/s$^{\beta}$). The flux memory thus decays algebraically with time: $D^{\ast}(t') \sim D_{\beta} \, t'^{-(1-\beta)}$.
\par
A conservation property emerges for fractional time superdiffusion in which the wait time distribution has a finite first moment $\left< \vartheta \right> = \tau$. These processes become diffusive in the long time limit with asymptotic diffusivity $D_{\infty} = L^2/\tau$. With $\Psi(s\rightarrow 0) \simeq s\tau$, we find
\begin{equation}
\mathbb{D}_t \geq 1 : D^{\ast}(s \rightarrow 0) = D_{\infty} \quad \leftrightarrow \quad \int \limits_{0}^{\infty} D^{\ast}(t') \mathrm{d}t' = D_{\infty} \label{Brownconservation}
\end{equation}
This forms a time-space duality with (\ref{Poissonconservation}): \textit{the total FGR temporal memory content in fractional time superdiffusion with long-term recovery is given by the asymptotic bulk diffusivity.} A specific example is given by tempered fractional time diffusion which we will investigate in detail in section IV.C.
\section{CASE STUDIES}
\subsection{Spatiotemporal flux memory in fractional diffusion equations}
Fractional diffusion equations (FDEs) describe system dynamics of the form
\begin{equation}
\frac{\partial^{\beta}}{\partial t^{\beta}} P(x,t) = D_{\alpha\beta} \cdot \frac{\partial^{\alpha}}{\partial |x|^{\alpha}} P(x,t) \qquad 0 < \alpha \leq 2 \,\, , \,\, 0 < \beta \leq 1 \label{FDE}
\end{equation}
with $\alpha$ and $\beta$ characteristic space and time exponents and $D_{\alpha\beta}$ a fractional diffusivity constant (unit m$^{\alpha}$/s$^{\beta}$). These expressions generalise the regular diffusion equation $\partial P/\partial t = D_0 \cdot \partial^2 P / \partial x^2$ to derivatives of non-integer order and as such were introduced as a natural candidate to model transport regimes with fractal space and/or time dimensions \cite{fractionalequations}.
\par
FDEs provide a valuable case study of memory effects for several reasons. First, they have been employed across several disciplines to describe a variety of anomalous dynamics \cite{klafter1,klafter2,fractionalequations}. In some applications, fractional differential operators even arise naturally from first principles \cite{FDEfirstprinciples}. Second, FDEs forge a connection between kinetic equations and the Montroll-Weiss stochastic framework \cite{fractionaldiffeq-CTRW}. In particular, it has been demonstrated \cite{fractionaldiffeq-CTRWasymptotics} that the solutions of the FDE describe the long-term dynamics of random processes in which jump length and wait time distributions follow L\'evy and Mittag-Leffler asymptotics respectively: $\psi(\xi \rightarrow 0) \sim |L \xi|^{\alpha}$ and $\Psi(s \rightarrow 0) \sim (s \tau)^{\beta}$ where $L$ and $\tau$ denote characteristic length and time scales as usual with $L^{\alpha}/\tau^{\beta} = D_{\alpha \beta}$. Finally, the two characteristic exponents personify the fractal properties of the system: $(\mathbb{D}_x,\mathbb{D}_t) = (\alpha,\beta)$. The derivation that follows will explain the necessary appearance of the additional differential operator(s) in the universal FGR postulated earlier in (\ref{convolution2}) when $\mathbb{D}_x < 1$ and/or $\mathbb{D}_t < 1$.
\par
Several types of fractional integro-differentiation operators, each mathematically self-consistent in their own right, have been developed \cite{fractional-textbook}. In our case, $\partial^{\beta}/\partial t^{\beta}$ denotes the Riemann-Liouville (RV) operator $_0 \mathcal{D}_t^{\beta}$ of order $\beta$ with lower bound zero. The operator is additive, in the sense $_0 \mathcal{D}_t^{\beta_1} \, {}_0 \mathcal{D}_t^{\beta_2} = {}_0 \mathcal{D}_t^{\beta_1+\beta_2}$ for arbitrary orders $\beta_1$ and $\beta_2$ \cite{fractionalequations}. We also have $_0 \mathcal{D}_t^{-\beta} = {}_0 \mathcal{I}_t^{\beta}$, where the fractional integration operator is defined as 
\begin{equation}
0 < \beta < 1 : \quad {}_0 \mathcal{I}_t^{\beta} \left[ g(t) \right] = \frac{1}{\Gamma(\beta)} \int \limits_{0}^{t} \frac{g(t') \mathrm{d}t'}{(t-t')^{1-\beta}} \label{fracint}
\end{equation}
The Laplace image of the operators satisfies $\mathcal{L}[{}_0 \mathcal{D}_t^{\pm \beta} g(t)] = s^{\pm \beta} \, G(s)$. In the same spirit, $\partial^{\alpha}/\partial |x|^{\alpha}$ is a symmetrised spatial differentiation operator of fractional order $\alpha$ defined in such a way that its Fourier image obeys $\mathcal{F}[\mathcal{D}_{|x|}^{\alpha} g(x)] = -|\xi|^{\alpha} \, G(\xi)$ \cite{klafter2,fractionalequations}. This prerequisite is clearly driven by physical motives, as it ensures that the resulting FDE solutions $P(x,t)$ are always non-negative and even in $x$, as appropriate for isotropic media. Mathematically, $\mathcal{D}_{|x|}^{\alpha}$ can be decomposed as a linear combination of two conventional RL operators $_{-\infty} \mathcal{D}_x^{\alpha}$ and $_{x} \mathcal{D}_{\infty}^{\alpha}$ \cite{fractionalequations,RL-2sided}.
\par
Accounting for initial conditions through source functions is slightly more involved in FDEs than in regular differential equations due to some peculiarities of fractional RV operators. Careful analysis \cite{klafter1,fractionalequations,fractionaltimediffusion} shows that the single pulse response of (\ref{FDE}), i.e. the solution that satisfies $P(x,t=0) = \delta(x)$, is determined by
\begin{equation}
\frac{\partial^{\beta}}{\partial t^{\beta}} P(x,t) = D_{\alpha\beta} \cdot \frac{\partial^{\alpha}}{\partial |x|^{\alpha}} P(x,t) + \frac{\delta(x) \, t^{-\beta}}{\Gamma(1-\beta)} \label{FDEsource}
\end{equation}
After Fourier-Laplace transform the solution reads
\begin{equation}
P(\xi,s) = \frac{s^{\beta - 1}}{s^{\beta} + D_{\alpha\beta} |\xi|^{\alpha}} \label{FDEPxis}
\end{equation}
From (\ref{FDEPxis}) it is easy to see that when $\beta = 1$, the process is Poissonian and the FDE describes L\'evy superdiffusion with fractal space dimension $\alpha$ which we discussed in section III.A. Conversely, when $\alpha = 2$ the kinetics reduce to Mittag-Leffler subdiffusion with characteristic time exponent $\beta$ as analysed in section III.B.
\par
Regarding the flux memory, a pure convolution FGR would produce
\begin{equation}
\text{if Eq. (\ref{convolution}) were valid:} \quad D^{\ast}(\xi,s) = \frac{D_{\alpha\beta}}{|\xi|^{2-\alpha} \, s^{\beta-1}}
\end{equation}
When $\alpha<1$, $D^{\ast}(\xi,s)$ is non-integrable near $\xi = 0$, and consequently its Fourier inversion diverges. Likewise, $D^{\ast}(\xi,s)$ exhibits a positive $s$ exponent when $\beta<1$, and has therefore no inverse Laplace transform. Given that both $P$ and $q$ always have meaningful Fourier-Laplace transforms due to physical constraints, the anomalies in $D^{\ast}(\xi,s)$ indicate a failure of the initial assumption (\ref{convolution}) itself. To derive an adequate form of the FGR, we start by applying $_0 \mathcal{D}_t^{1-\beta}$ to both sides of (\ref{FDEsource}). Operator additivity and conservation of energy/mass yields
\begin{equation}
x \neq 0 : \quad \frac{\partial}{\partial t} P(x,t) = -\frac{\partial}{\partial x} q(x,t) = D_{\alpha\beta} \cdot \frac{\partial}{\partial t} \left[ {}_0 \mathcal{I}_{t}^{\beta} \frac{\partial^{\alpha}}{\partial |x|^{\alpha}}P(x,t) \right] \label{FGRderivation}
\end{equation}
A distinction regarding $\alpha$ is needed to proceed further.
\par
For $1 < \alpha < 2$, one can show (full proof in Appendix A):
\begin{equation}
\frac{\partial^{\alpha}}{\partial |x|^{\alpha}} P(x,t) = \frac{1}{2 \Gamma (2-\alpha) \cos[(2-\alpha) \frac{\pi}{2}]} \cdot \frac{\partial}{\partial x} \int \limits_{-\infty}^{\infty} \frac{\frac{\partial P}{\partial x}(x',t) \mathrm{d}x'}{|x-x'|^{\alpha-1}}
\end{equation}
Inserting this into (\ref{FGRderivation}) leads to an equation of the form $-\partial_x q(x,t) = \partial_x K(x,t)$ and thus $q(x,t) = -K(x,t) + A$. The integration constant $A$ is subsequently found to be zero since both flux and gradient must physically vanish at $x \rightarrow \pm \infty$. After applying (\ref{fracint}) we have
\begin{equation}
q(x,t) = - \frac{D_{\alpha\beta}}{2 \Gamma (2-\alpha) \Gamma(\beta) \cos[(2-\alpha) \frac{\pi}{2}]} \cdot \frac{\partial}{\partial t} \int \limits_{0}^{t} \frac{\mathrm{d}t'}{(t-t')^{1-\beta}} \int \limits_{-\infty}^{\infty} \frac{\mathrm{d}x'}{|x-x'|^{\alpha-1}} \frac{\partial P}{\partial x}(x',t')
\end{equation}
which is of the form postulated in (\ref{convolution2}) with $(\chi,\eta) = (0,1)$ and
\begin{equation}
D^{\ast}(x',t') = \frac{D_{\alpha\beta} \, |x'|^{-(\alpha-1)} \, t'^{-(1-\beta)}}{2 \Gamma (2-\alpha) \Gamma(\beta) \cos[(2-\alpha) \frac{\pi}{2}]} \label{Dstar01}
\end{equation}
The FGR memory decays algebraically in both space and time. A result identical to (\ref{Dstar01}) is found directly through Fourier-Laplace inversion of $D^{\ast}(\xi,s) = D_{\alpha\beta}/s^{\beta}|\xi|^{2-\alpha}$ given by (\ref{kappamacro}).
\par
For $0<\alpha<1$, we find (full proof in Appendix B):
\begin{equation}
\frac{\partial^{\alpha}}{\partial |x|^{\alpha}} P(x,t) = \frac{-1}{2 \Gamma (1-\alpha) \sin[(1-\alpha) \frac{\pi}{2}]} \cdot \int \limits_{-\infty}^{\infty} \frac{\mathrm{sgn}(x-x') \frac{\partial P}{\partial x}(x',t) \mathrm{d}x'}{|x-x'|^{\alpha}}
\end{equation}
with $\mathrm{sgn}(\cdot)$ the sign function. Insertion into (\ref{FGRderivation}) and application  of (\ref{fracint}) yields
\begin{equation}
\frac{\partial}{\partial x} q(x,t) = \frac{D_{\alpha\beta}}{2 \Gamma (1-\alpha) \Gamma(\beta) \sin[(1-\alpha) \frac{\pi}{2}]} \cdot \frac{\partial}{\partial t} \int \limits_{0}^{t} \frac{\mathrm{d}t'}{(t-t')^{1-\beta}} \int \limits_{-\infty}^{\infty} \frac{\mathrm{sgn}(x-x') \mathrm{d}x'}{|x-x'|^{\alpha}} \frac{\partial P}{\partial x}(x',t')
\end{equation}
This agrees with the form (\ref{convolution2}) when setting $(\chi,\eta) = (1,1)$ with
\begin{equation}
D^{\ast}(x',t') = - \frac{D_{\alpha\beta} \, \mathrm{sgn}(x') |x'|^{-\alpha} \, t'^{-(1-\beta)}}{2 \Gamma (1-\alpha) \Gamma(\beta) \sin[(1-\alpha) \frac{\pi}{2}]}
\end{equation}
The same result is found from inversion of $D^{\ast}(\xi,s) = j D_{\alpha\beta} \, \mathrm{sgn}(\xi)/s^{\beta}|\xi|^{1-\alpha}$ provided by (\ref{kappamacro}). Once again we find an algebraic decay in both space and time. Notice however that here the kernel is odd with respect to space.
\subsection{Spatial flux memory in tempered L\'evy transport}
A variety of disciplines have observed Poissonian dynamics that exhibit a gradual transition from a superdiffusive L\'evy regime to regular diffusion. Examples include sediment transport in rivers \cite{geophysics-temperedlevy,hydrology-temperedlevy}, the evolution of financial markets \cite{finance-temperedlevy1,finance-temperedlevy2}, and quasiballistic heat conduction in semiconductor alloys \cite{alloys-part1,alloys-part2}. Such behaviour can be described by so called tempered or truncated L\'evy theory \cite{truncatedlevy1,truncatedlevy2}. One possibility is to accelerate the algebraic dacay of the pure L\'evy jump length distribution with an exponential in order to suppress the likelihood of extremely long jumps: $\phi(\zeta) \sim \exp(-|\zeta|/\zeta_0)/|\zeta|^{1+\alpha}$ \cite{truncatedlevy2,alloys-part2}. Here, we will use a simplified approach that still preserves all essential trends:
\begin{equation}
D^{\ast}(\xi) = \frac{\psi(\xi)}{\tau \, \xi^2} = \frac{D_{\infty}}{\left( 1 + x_{\text{R}}^2 \xi^2 \right)^{1-\alpha/2}} \quad 1 < \alpha < 2 \label{DstarLF}
\end{equation}
One can easily verify this corresponds to L\'evy transport with fractal space dimension $\alpha$ and $D_{\alpha} = D_{\infty}/x_{\text{R}}^{2-\alpha}$ that recovers to regular diffusion over characteristic length and time scales $x_{\text{R}}$ and $t_{\text{R}} = x_{\text{R}}^2/D_{\infty}$ respectively. A key advantage is that (\ref{DstarLF}) can be transformed analytically to real space domain, enabling detailed study of its shape:
\begin{equation}
D^{\ast}(x') = \frac{2^{\nu} D_{\infty}}{\sqrt{\pi} \, \Gamma (1/2-\nu) \, x_{\text{R}}} \cdot \frac{K_{\nu} (|x'/x_{\text{R}}|)}{|x'/x_{\text{R}}|^{\nu}} \quad \quad \nu = \frac{\alpha-1}{2} \label{kappaLF}
\end{equation}
with $K$ the modified Bessel function of the second kind. The short-range core of the spatial memory kernel exhibits pure L\'evy behaviour $D^{\ast}(|x'| \ll x_{\text{R}}) \sim |x'/x_{\text{R}}|^{-(\alpha-1)}$, while the tails decay much more rapidly as $|x'/x_{\text{R}}|^{-\alpha/2} \exp (- |x'/x_{\text{R}}|)$. Since the tempered L\'evy process obeys the conservation property (\ref{Poissonconservation}), the total area under $D^{\ast}(x')$ always equals $D_{\infty}$. The tempered L\'evy parameters determine how this fixed memory budget gets spatially distributed: $x_{\text{R}}$ sets the overall length scale while $\alpha$ regulates the shape (Fig. 1a).
\myfig[!htb]{width=\textwidth}{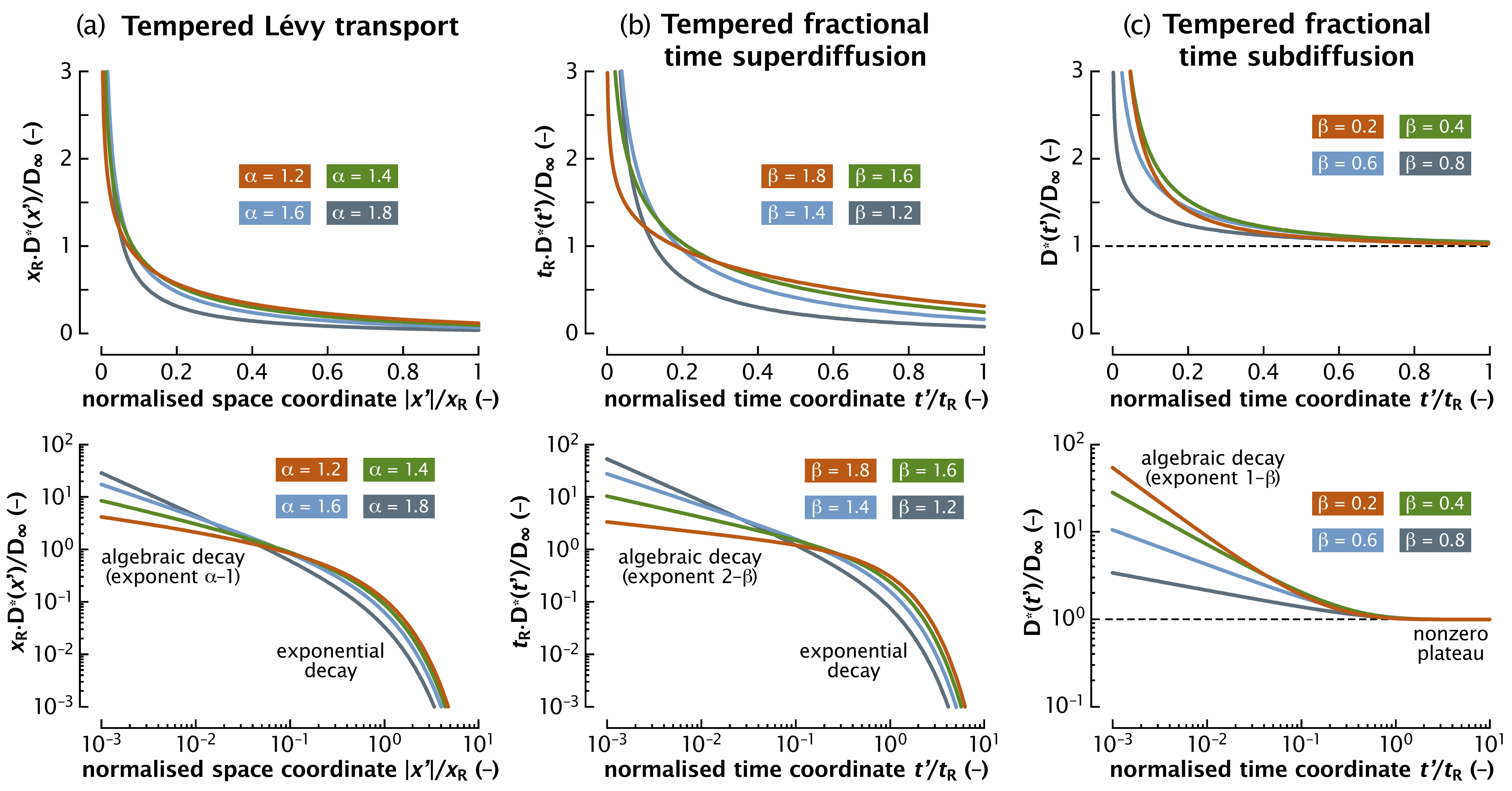}{Normalised flux memory in commonly encountered types of anomalous transport with long-term diffusive recovery (asymptotic bulk diffusivity $D_{\infty}$). (a) Spatial flux memory in tempered L\'evy superdiffusion as given by analytical expression (\ref{kappaLF}). $\alpha$ is the fractal space dimension of the L\'evy regime while $x_{\text{R}}$ determines the characteristic length scale over which diffusive recovery occurs. (b) Temporal flux memory in fractional time superdiffusion as given by analytical expression (\ref{Dstarsuperdiffusion}). $\beta$ is the fractal time dimension of the superdiffusion regime and $t_{\text{R}}$ determines the characteristic time scale of the diffusive recovery. (c) Temporal flux memory in fractional time subdiffusion obained by numerical Laplace inversion of (\ref{Dstartemperedfractional}), with $\beta$ and $t_{\text{R}}$ analogous to before.}
\par
As $\alpha$ decreases under constant $x_{\text{R}}$, $D^{\ast}$ becomes less sharply concentrated near the origin in favour of more prominent tails. This qualitatively illustrates that the FGR delocalisation becomes more pronounced as the transport dynamics deviate more severely from regular diffusion, as intuitively expected. To quantify the length scale over which nonlocal effects are physically important, we exploit that $D^{\ast}(x')/D_0$ can be treated mathematically as a properly normalised probability density. This results in a well defined standard deviation that can be calculated in closed form: $\sqrt{\left< X'^2 \right>} = \sqrt{2-\alpha} \,\, x_{\text{R}}$. Therefore, \textit{the effective extent of the spatial flux memory is on the order of the diffusive recovery length of the macroscopic field}. In the diffusive limit $\alpha \rightarrow 2$, the characteristic width of the kernel goes to zero, signaling that the spatial memory collapses into a single Dirac peak as appropriate.
\par
These insights remain hidden in conventional methodologies that solely rely on the tracer density $P(x,t)$ and flux-gradient diagrams (Fig. 2).
\myfig[!htb]{width=\textwidth}{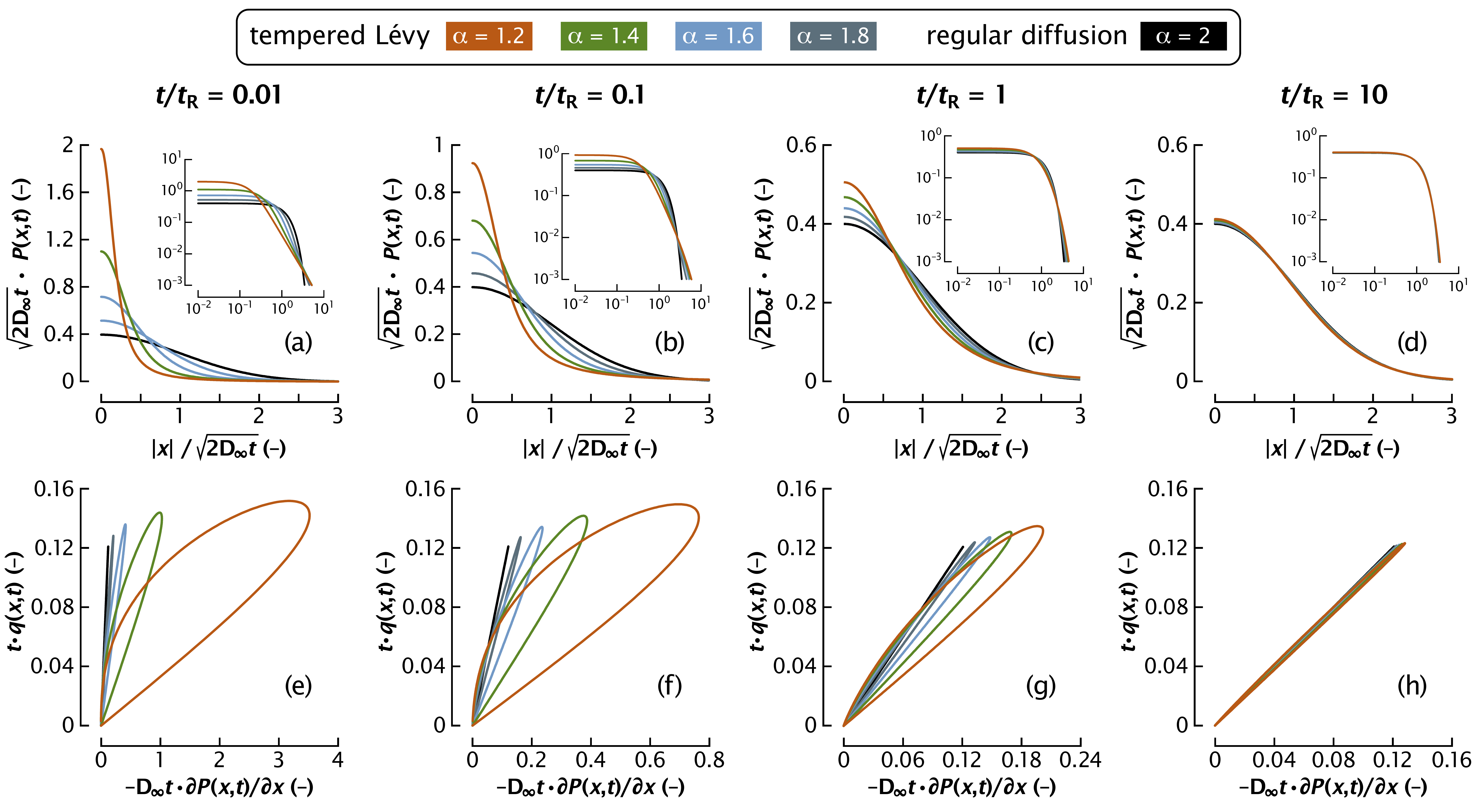}{Tempered Poissonian L\'evy transport governed by jump length dynamics (\ref{kappaLF}): (a)--(d) Normalised tracer distributions at various times (insets show graphs on double logarithmic scale); (e)--(h) Normalised flux-gradient diagrams at various times. The pure L\'evy regime with fractal space dimension $\alpha$ asymptotically recovers over characteristic length scale $x_{\text{R}}$ and associated time scale $t_{\text{R}} = x_{\text{R}}^2 / D_{\infty}$ to regular diffusion with bulk diffusivity $D_{\infty}$. All curves are obtained by numerical evaluation of (\ref{Pxtpoisson}) and (\ref{qxtpoisson}).}
\par
The anomalous character of the process and gradual diffusive recovery over time scales $\simeq t_{\text{R}}$ are obvious. However, from these diagrams it is not at all clear that the core dynamics are governed by a purely spatial FGR delocalisation that applies at all times, and much less how the shape of the associated memory kernel looks like.
\par
The practical capabilities of our nonlocal formalism can be illustrated by the experimental analysis of microscale heat conduction in semiconductors. A detailed study will be presented elsewhere \cite{microscaleheatconduction}. Briefly, the methodology relies on expressions similar to (\ref{kappaLF}) to model the transition from ballistic to diffusive thermal transport regimes. From $D^{\ast}(\xi)$ it is then possible to derive closed form predictions for the effective thermal conductivity inferred by transient thermal grating and time domain thermoreflectance experiments that closely match actual measurement data. The extent of the spatial heat flux memory is found to be about half a micron in single crystals and two to three microns in alloy materials, in good agreement with the median mean free paths of the microscopic heat carriers.
\subsection{Temporal flux memory in tempered fractional time diffusion}
In time-space analogy to the previous section, we investigate tempered versions of fractional time diffusion by subjecting the regular algebraic carrier wait time distribution to an exponential tail. This could be achieved directly by choosing a gamma distribution for $\varphi(\vartheta)$. Instead we will again use a slightly simpler approach that captures the same trends but enables some analytical treatments, namely
\begin{equation}
\frac{\Psi(s)}{L^2} = \frac{s}{D_{\infty} (1+s t_{\text{R}})^{1-\beta}} \quad 0 < \beta < 2 \,\, , \,\, t_{\text{R}} > 0 \label{Psitemperedfractional}
\end{equation}
One can verify that the associated wait time distribution indeed transitions from an algebraic relation $\varphi(\vartheta \ll \tau) \sim (\vartheta/\tau)^{\beta-1}$ to an exponential decay $\varphi(\vartheta \gg \tau) \sim \exp(-\vartheta/\tau)$ where $\tau = L^2/D_{\infty}$. Note that $\beta=1$ corresponds to the simple Poissonian case $\Psi(s) = s \tau$ discussed earlier. From (\ref{PBrownian}) we see that this process indeed describes the desired transition:
\begin{eqnarray}
s t_{\text{R}} \gg 1 & : & P(\xi,s) \simeq \frac{s^{\beta - 1}}{s^{\beta} + D_{\beta} \xi^2} \\
s t_{\text{R}} \ll 1 & : & P(\xi,s) \simeq \frac{1}{s + D_{\infty} \xi^2}
\end{eqnarray}
At early times, the process is governed by fractional time diffusion with characteristic exponent $\beta$ and fractional diffusivity $D_{\beta} = D_{\infty} t_{\text{R}}^{1-\beta}$, with MSD $\sigma^2(t \ll t_{\text{R}}) \simeq 2 D_{\beta} t^{\beta}/\Gamma(\beta+1)$. At long times, the process becomes diffusive with bulk diffusivity $D_{\infty}$. The cross-over between the two asymptotic regimes occurs over time scales on the order of $t_{\text{R}}$. The temporal FGR memory of the considered process is given by
\begin{equation}
D^{\ast}(s) = \frac{D_{\infty} (1+s t_{\text{R}})^{1-\beta}}{s^{\eta}} \label{Dstartemperedfractional}
\end{equation}
A distinction regarding $\beta$ is useful for detailed investigation.
\subsubsection{Tempered fractional time superdiffusion ($1 < \beta < 2$, $\eta = 0$)}
As $\beta$ increases beyond unity, the short time transport dynamics become more and more deterministic, since in the limit $\beta \rightarrow 2$ the associated FDE tends to the wave equation with propagation velocity $\sqrt{D_{\beta}}$. This leads to several notable features in the tracer distributions and flux-gradient diagrams (Fig. 3).
\myfig[!htb]{width=\textwidth}{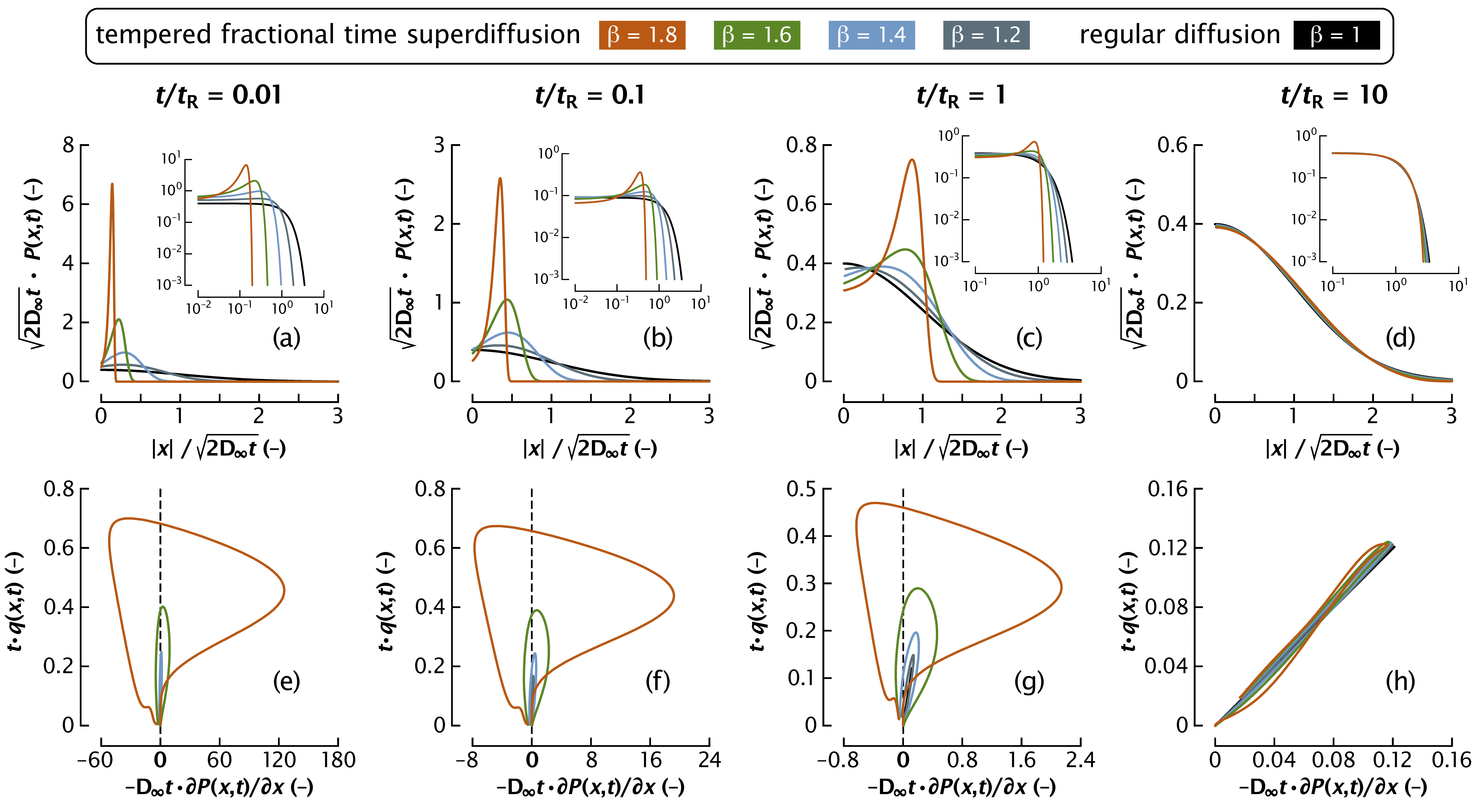}{Tempered fractional time superdiffusion governed by wait time dynamics (\ref{Psitemperedfractional}): (a)--(d) Normalised tracer distributions at various times (insets show graphs on double logarithmic scale); (e)--(h) Normalised flux-gradient diagrams at various times. The superdiffusive regime with fractal time dimension $\beta$ asymptotically recovers over characteristic time scale $t_{\text{R}}$ to regular diffusion with bulk diffusivity $D_{\infty}$. All curves are obtained by numerical evaluation of (\ref{PxtBrownian}) and (\ref{qxtBrownian}).}
\par
During the anomalous regime the tracer distributions are bimodal, meaning that the maximum tracer density occurs internally in the medium on either side of the source instead of at the source itself (Fig 3a--c). As $\beta$ increases, the peaks become progressively sharper and appear closer to the source. Meanwhile, calculations show that the flux still points away from the source at all places and times. This leads to `uphill' transport (flux and gradient have equal signs) in the region between the source and the distribution peak, as signified by the flux-gradient diagrams crossing into the second quadrant (Fig. 3e--g). We note that uphill transport has been reported earlier in flux-gradient diagrams of tempered L\'evy processes with skewed jump length distributions \cite{fractionalfick-nonlocal}. However, in that case the cross-gradient effect is due to the inherent directional bias caused by the probability imbalance between left and right jumps, while the process considered here is still fully symmetric.
\par
Based on the tracer distributions and flux-gradient diagrams, tempered fractional time superdiffusion behaves very differently from the earlier discussed tempered (fractional space) L\'evy transport (Fig. 2), despite the strong space-time duality in their internal process dynamics. By contrast, the inherent similarities become immediately obvious in the corresponding FGR memories. The $D^{\ast}(s)$ kernel given by (\ref{Dstartemperedfractional}) can be inverted analytically:
\begin{equation}
D^{\ast}(t') = \frac{D_{\infty}}{t_{\text{R}}} \cdot \frac{\exp(-t'/t_{\text{R}})}{\Gamma(\beta-1) \, (t'/t_{\text{R}})^{2-\beta}} \label{Dstarsuperdiffusion}
\end{equation}
The short-term memory decays algebraically while the tails are exponential, just like we observed for (\ref{kappaLF}). Plots of the temporal flux memory (Fig. 1b) indeed show a striking resemblance to the spatial flux memory in tempered L\'evy transport (Fig. 1a). Given that conservation property (\ref{Brownconservation}) applies, the time scale over which memory effects are physically important can by quantified by treating $D^{\ast}(t')/D_{\infty}$ as a probability distribution. Its mean and standard deviation are found to be $(\beta-1)\, t_{\text{R}}$ and $\sqrt{\beta-1} \,\, t_{\text{R}}$ respectively, and therefore \textit{the effective extent of the temporal flux memory is on the order of the diffusive recovery time scale of the macroscopic field}. This is once again the dual form of the result found earlier for tempered L\'evy processes. The characteristic width of the kernel goes to zero in the purely diffusive limit $\beta \rightarrow 1$, signaling appropriate collapse into the memoryless Fick/Fourier law.
\subsubsection{Tempered fractional time subdiffusion ($0 < \beta < 1$, $\eta = 1$)}
The gradual transition from (Mittag-Leffler) subdiffusion to regular diffusion is commonly observed in geophysical applications for the evolution of tracer plumes through heterogeneous media \cite{geophysics-temperedsubdiffusion1,geophysics-temperedsubdiffusion2}. Subdiffusive regimes have a characteristic cusp in the center of the tracer distribution (the gradient is nonzero at the source) which results in open flux-gradient diagrams that do not start from the origin (Fig. 4). The diffusive recovery also induces an interesting density overshoot near the source (Fig. 4c). 
\myfig[!htb]{width=\textwidth}{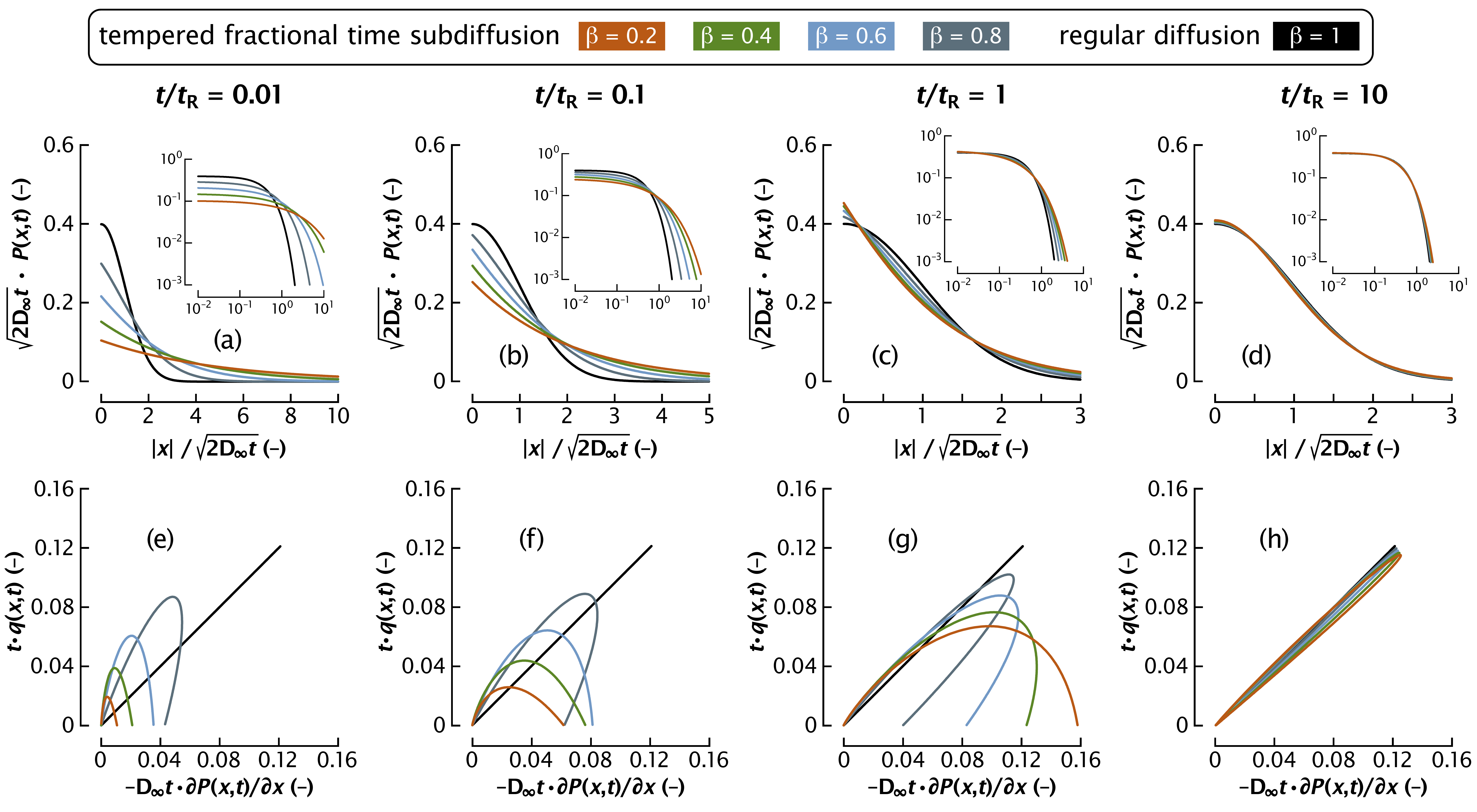}{Tempered fractional time subdiffusion governed by wait time dynamics (\ref{Psitemperedfractional}): (a)--(d) Normalised tracer distributions at various times (insets show graphs on double logarithmic scale); (e)--(h) Normalised flux-gradient diagrams at various times. The subdiffusive regime with fractal time dimension $\beta$ asymptotically recovers over characteristic time scale $t_{\text{R}}$ to regular diffusion with bulk diffusivity $D_{\infty}$. All curves are obtained by numerical evaluation of (\ref{PxtBrownian}) and (\ref{qxtBrownian}).}
\pagebreak[4]\par
The associated memory kernel must be evaluated from numerical Laplace inversion of $D^{\ast}(s)$, though we can still derive the asymptotics in closed form:
\begin{eqnarray}
t' \ll t_{\text{R}} & : & D^{\ast}(t') \simeq \frac{D_{\infty}}{\Gamma(\beta)} (t/t_{\text{R}})^{-(1-\beta)} \\
t' \gg t_{\text{R}} & : & D^{\ast}(t') \simeq D_{\infty}
\end{eqnarray}
The temporal memory first decays algebraically, and then levels out to the asymptotic bulk diffusivity (Fig. 1c). In the limit $\beta \xrightarrow{<} 1$, $D^{\ast}(t') \equiv D_{\infty}$ at all $t'$. This seems counterintuitive as one may expect a gradual sharpening towards a Dirac peak as the process becomes more and more Poissonian. However, this result stems from the presence of the $\partial_t$ operator in (\ref{convolution2}) for $\eta = 1$. As $D^{\ast} \rightarrow D_{\infty}$, the convolution reduces to a simple time integral of the gradient with constant scaling factor. The $\partial_t$ operator then differentiates this result again, so we recover the conventional (localised) FGR $q(x,t) = -D_{\infty} \, \partial_x P(x,t)$ as appropriate.
\section{Conclusions}
In summary, we developed a universal formalism that characterises the inherent nonlocality in anomalous transport processes. A generalised diffusivity kernel $D^{\ast}$ fully embodies the spatial and temporal memory of the tracer flux with respect to the tracer density gradient. We obtained analytical expressions for the shape and physical extent of the flux memory in several commonly encountered types of nondiffusive transport. The framework also clearly conveys fundamental space-time dualities in the underlying dynamics that remain completely hidden in conventional analyses based on the tracer density distribution and flux-gradient diagrams. Practical capabilities are illustrated by, but certainly not limited to, experimental characterisations of the spatial flux memory in microscale heat superdiffusion.
\section*{ACKNOWLEDGEMENTS}
B.V. is grateful to Prof. Philip Allen for discussions that inspired and motivated this work. A.S. acknowledges Purdue University Start Up funds that supported B.V.'s basic research.
\appendix
\section{Symmetrised space derivative of fractional order between 1 and 2}
Let $g(x)$ be an integrable function over the entire real axis. We seek to write its symmetrised fractional derivative of order $1 < \alpha < 2$ as
\begin{equation}
\frac{\partial^{\alpha} g}{\partial |x|^{\alpha}} = \frac{\partial}{\partial x} \left[ w_{\alpha}(x) \ast \frac{\partial g}{\partial x} \right] \label{appfrac1}
\end{equation}
where $\ast$ denotes convolution and $w_{\alpha}(x)$ is an unknown kernel function to be determined. Fourier transformation provides
\begin{equation}
-|\xi|^{\alpha} G(\xi) = j\xi [W_{\alpha}(\xi) \cdot j \xi G(\xi)]
\end{equation}
where the left hand side arises \textit{by construction}, since the operator $\partial^{\alpha}/\partial |x|^{\alpha}$ \textit{is defined with precisely this property in mind}. $G(\xi)$ cancels out, showing that (\ref{appfrac1}) is a viable form for the fractional derivative, and we find
\begin{equation} 
W_{\alpha}(\xi) = \frac{1}{|\xi|^{2-\alpha}}
\end{equation}
Fourier inversion yields
\begin{equation}
w_{\alpha}(x) = \frac{1}{2\pi} \int \limits_{-\infty}^{\infty} \frac{\exp(j \xi x) \mathrm{d}\xi}{|\xi|^{2-\alpha}} = \frac{1}{\pi} \int \limits_{0}^{\infty} \frac{\cos(\xi x) \mathrm{d}\xi}{\xi^{2-\alpha}} = \frac{|x|^{-(\alpha-1)}}{2 \Gamma (2-\alpha) \cos[(2-\alpha) \frac{\pi}{2}]}
\end{equation}
Therefore (\ref{appfrac1}) becomes
\begin{equation}
1 < \alpha < 2 : \quad \frac{\partial^{\alpha}g}{\partial |x|^{\alpha}} = \frac{1}{2 \Gamma (2-\alpha) \cos[(2-\alpha) \frac{\pi}{2}]} \cdot \frac{\partial}{\partial x} \int \limits_{-\infty}^{\infty} \frac{\frac{\partial g}{\partial x}(x') \mathrm{d}x'}{|x-x'|^{\alpha-1}}
\end{equation}
as employed in the main text.
\section{Symmetrised space derivative of fractional order between 0 and 1}
Let $g(x)$ be an integrable function over the entire real axis. We seek to write its symmetrised fractional derivative of order $0 < \alpha < 1$ as
\begin{equation}
\frac{\partial^{\alpha} g}{\partial |x|^{\alpha}} = w_{\alpha}(x) \ast \frac{\partial g}{\partial x} \label{appfrac2}
\end{equation}
In a similar fashion as above, we find that the Fourier image of the kernel function $w_{\alpha}(x)$ is given by
\begin{equation} 
W_{\alpha}(\xi) = \frac{j \cdot \mathrm{sgn}(\xi)}{|\xi|^{1-\alpha}}
\end{equation}
Fourier inversion yields
\begin{equation}
w_{\alpha}(x) = \frac{1}{2\pi} \int \limits_{-\infty}^{\infty} \frac{j \cdot \mathrm{sgn}(\xi) \exp(j \xi x) \mathrm{d}\xi}{|\xi|^{1-\alpha}} = -\frac{1}{\pi} \int \limits_{0}^{\infty} \frac{\sin(\xi x) \mathrm{d}\xi}{\xi^{1-\alpha}} = \frac{-\mathrm{sgn}(x) \cdot |x|^{-\alpha}}{2 \Gamma (1-\alpha) \sin[(1-\alpha) \frac{\pi}{2}]}
\end{equation}
Now (\ref{appfrac2}) becomes
\begin{equation}
0 < \alpha < 1 : \quad \frac{\partial^{\alpha}g}{\partial |x|^{\alpha}} = \frac{-1}{2 \Gamma (1-\alpha) \sin[(1-\alpha) \frac{\pi}{2}]} \cdot \int \limits_{-\infty}^{\infty} \frac{ \mathrm{sgn}(x-x') \cdot \frac{\partial g}{\partial x}(x') \mathrm{d}x'}{|x-x'|^{\alpha}}
\end{equation}
as employed in the main text.
%

\end{document}